# Triple Higgs couplings at LHC

Alain Le Yaouanc[1], François Richard[2]

Université Paris-Saclay, CNRS/IN2P3, IJCLab, 91405 Orsay, France

July 2024

**Abstract**

*A global interpretation of several significant indications of scalar resonances observed at LHC is achieved using the* ***Georgi Machacek model GM****. Among many other consequences, one predicts large cross sections for the processes ggF→H(320)→H(125)H(125) and A(151)A(151) and ggF→A(420)→H(320)Z, where* ***H(320****) has been observed in A(420)→H(320)Z→bbbbℓ+ℓ- and where* ***A(151)*** *is observed into two photons accompanied by a b jet, a lepton or missing transverse energy. We predict that ggF→H(320) has a cross section of about 2000 fb with a BR into H(125)H(125) of 17% and into A(151)A(151) of 45%. Henceforth we predict that ggF→H(320)→HH will dominate over the* ***SM process H*→HH,*** *where H is the SM H(125) scalar. One expects that* ***H(320)→A(151)A(151)*** *can be observed into bbbb, becoming one of the most significant* ***BSM*** *phenomena identified so far. Arguments in favor of a SUSY version of GM are also provided.*

**Work for the 2024 International Workshop on Future Linear Colliders (LCWS2024)**

1 Alain Le Yaouanc <alain.le-yaouanc@ijclab.in2p3.fr>
2 François Richard <francois.richard@ijclab.in2p3.fr>

# I. Introduction

Measuring the triple Higgs coupling H*→HH has been identified as a major goal of our field, our 'holy grail', a direct exploration of the Higgs potential. The SM model process has a very small cross section and can be affected if there are resonances decaying into HH. In this note we will describe a scenario where there is such a resonance and how it can be discovered with an optimized strategy. We will show that this resonance provides an interpretation of the pseudo-scalar A(151), which is seen into two photons with an additional activity such as extra b jets, leptons and missing energy.

Recall that several indications for light scalars have been observed in a recent past. Relying on the **Georgi Machacek** model, GM, and on **Haber et al. sum rules**, an attempt for a global interpretation has been presented in [1].

Here we focus on a promising aspect of this work concerning the scalar **H(320)** observed in the cascade **A(420)→H(320)Z→bbbbℓ+ℓ-** [2], interpreted as coming from **H(320)→H(125)H(125)**. This result is recalled and interpreted in the appendix. This mechanism constitutes a **source of bbbb events** which can be tagged using the Z decays into leptons and neutrinos. [2] provides an upper limit of about 70 fb for this process assuming that only HH contributes.

Interpreting H(320) as the companion of H+(375) and H++(450) in a five-plet representation of the GM model, we find that ggF→H(320) has a cross of 2000 fb. This interpretation relies on our ability to predict the coupling $g_{H(320)tt}$ using the **'matrix method'** described in [1]. Our result says that the **Yukawa coupling** $Y_{H(320)tt}$ , normalized to the SM value, is ~ 0.52.

GM predicts that H(320) will decay into A(151)A(151) with a BR of 45%. We interpret the origin of the various signatures accompanying the scalar **A(151)** observed into 2$\gamma$ [3] as coming either from the decays into bb and $\tau\tau$ of the spectator A(151) or from Z decays in the process A(420)->H(320)Z.

There are therefore two sources of A(151)A(151) (and H(125)H(125)):

- One originating from the cascade A(420)→H(320)Z→A(151)A(151)Z which provides tagging of the 2$\gamma$ decays either from A(131)→bb, $\tau\tau$ or from Z→bb,ℓ+ℓ-,$\nu\nu$
- The other originating from ggF→A(151)A(151) with tagging from bb, $\tau\tau$

We predict that H(320) will decay into **HH** with a BR of 17.5%, the later result showing that the present SM searches for di-Higgs will be dominated by the contribution of this resonance, a **critical result**.

Given its large coupling to scalars, H(320) is very wide, $\Gamma$tot=220 GeV. One also predicts $\Gamma_{ZZ}$=5 GeV, meaning that with such a small BR, this mode has not been detected at LHC.

# II. Impact of H(320) on the searches for HH → 4b

Let us first recall that present searches for HH combine final states like $\gamma\gamma$/WW/bb/$\tau\tau$. The 4b final state has the highest BR but with a large QCD background.

The following bi-dimensional plot [5] indicates the distribution of the background for the observed bb sub-masses. This background peaks closely to the region of the HH signal region, defined by the SR contour, rendering the detection challenging. This is even more the case for A(151)A(151) which will be centered on the maximum background but, as we will see below, with an expected cross section two orders of magnitude larger than the SM process.

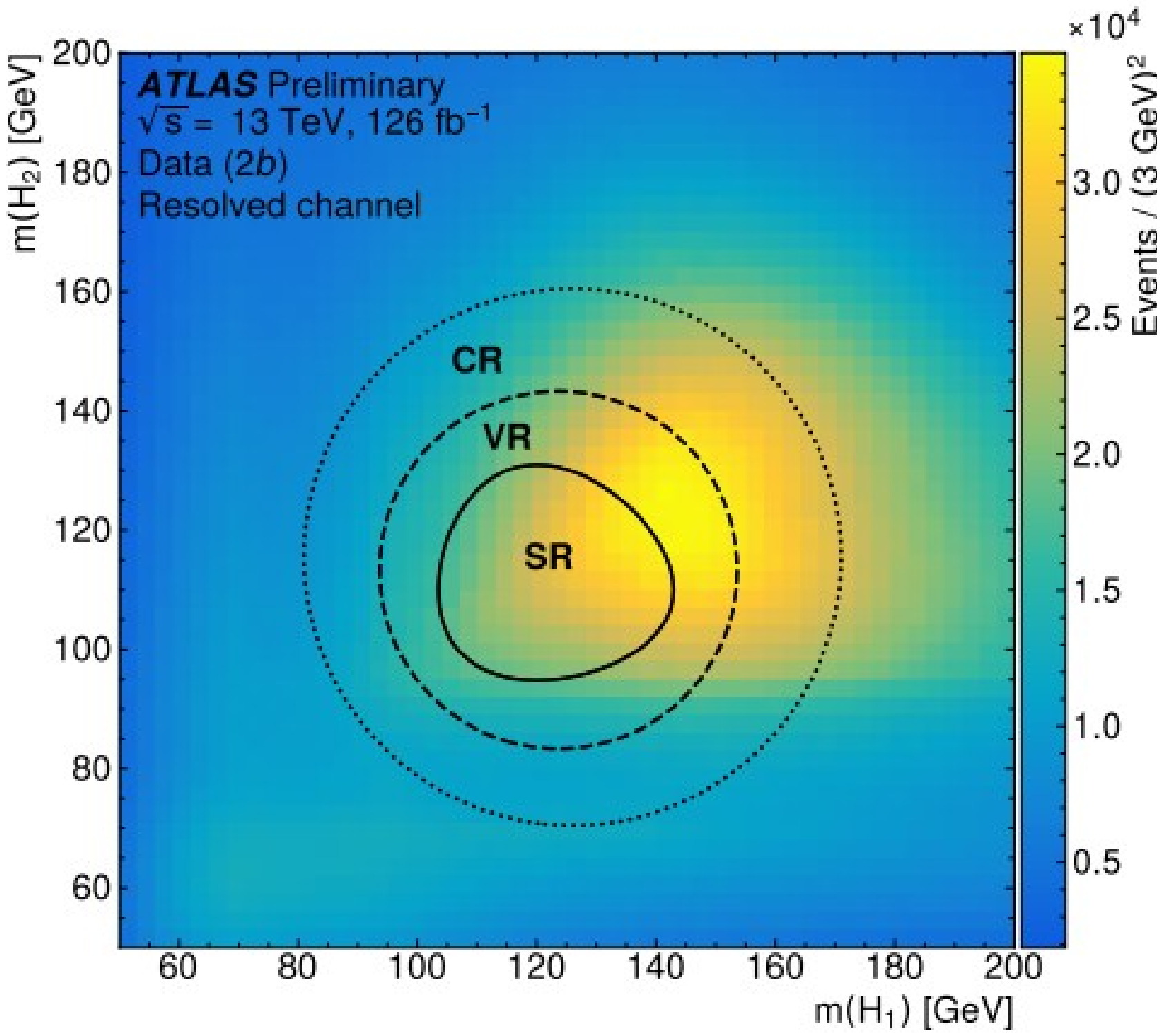

*Figure 1: mH1 and mH2 are the reconstructed masses of the Higgs bosons candidates in 4b final states, sorted by pT.*

From this figure one may infer that the SR selection only keeps about 1% (crude estimate) of the A(151)A(151) population which explains that in spite of its much higher rate, there is no significant excess with respect to the background as shown in figure 2 describing the limit achieved by ATLAS [5].

For the SM, one expects σggF->HH = 31 fb for $\kappa_\lambda$ ~1. Taking into account the BR into bb, one expects a 10 fb cross section in 4b. For low HH masses the efficiency in 4 b is of order 1%, meaning a negligible amount of events with respect to the huge QCD background.

Within GM, one may expect [1] significant deviations of $\kappa_\lambda$ from the SM. Figure 3 indicates how the m(HH) distribution varies, showing that for the predicted value [1] $\kappa_\lambda$ ~5, this distribution peaks in the region of the H(320) resonance but, again, with a much smaller cross section.

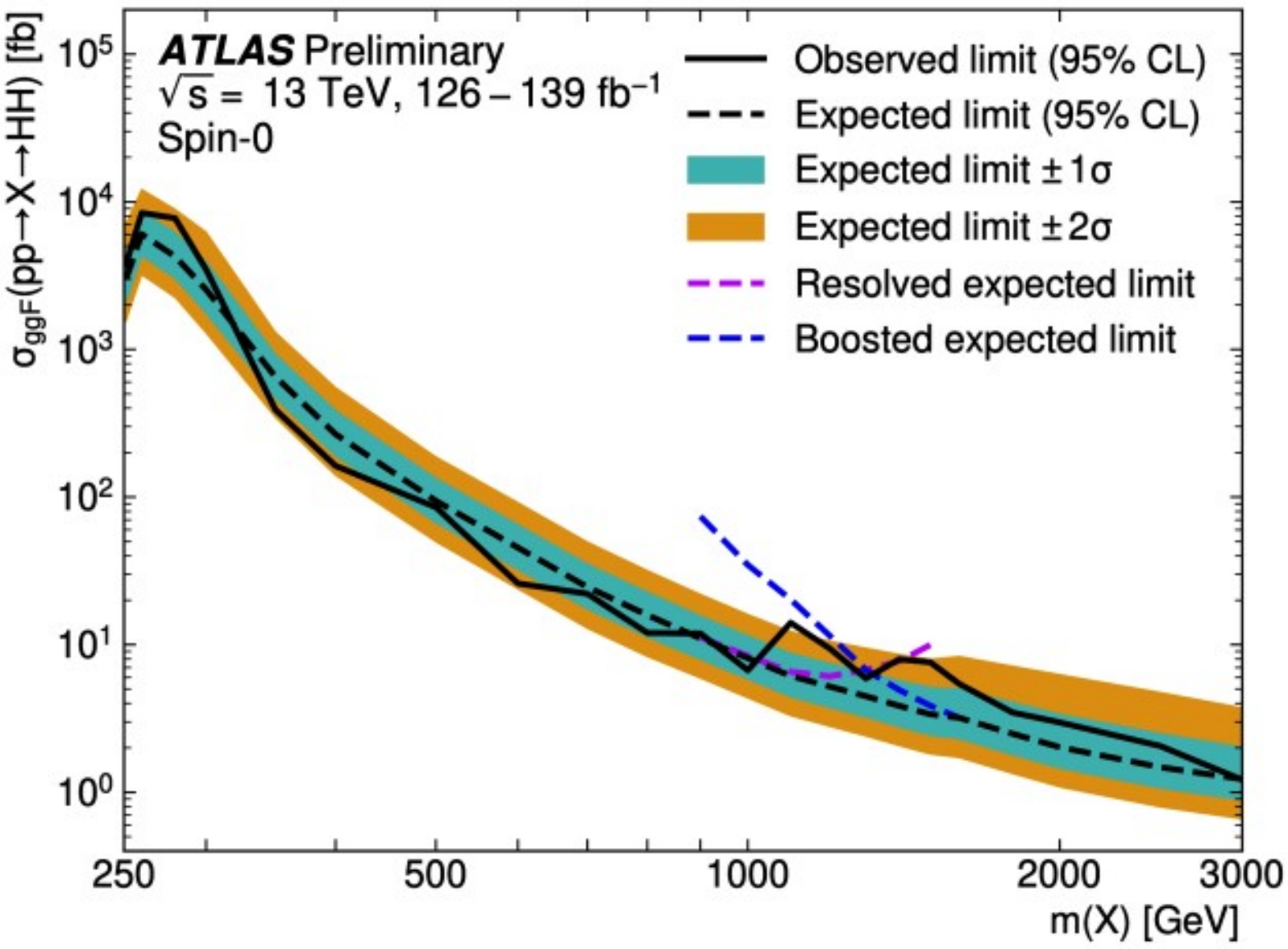

*Figure 2: Expected (dashed black lines) and observed (solid black lines) 95% CL upper limits on the cross section times branching ratio of spin zero resonant X→HH production.*

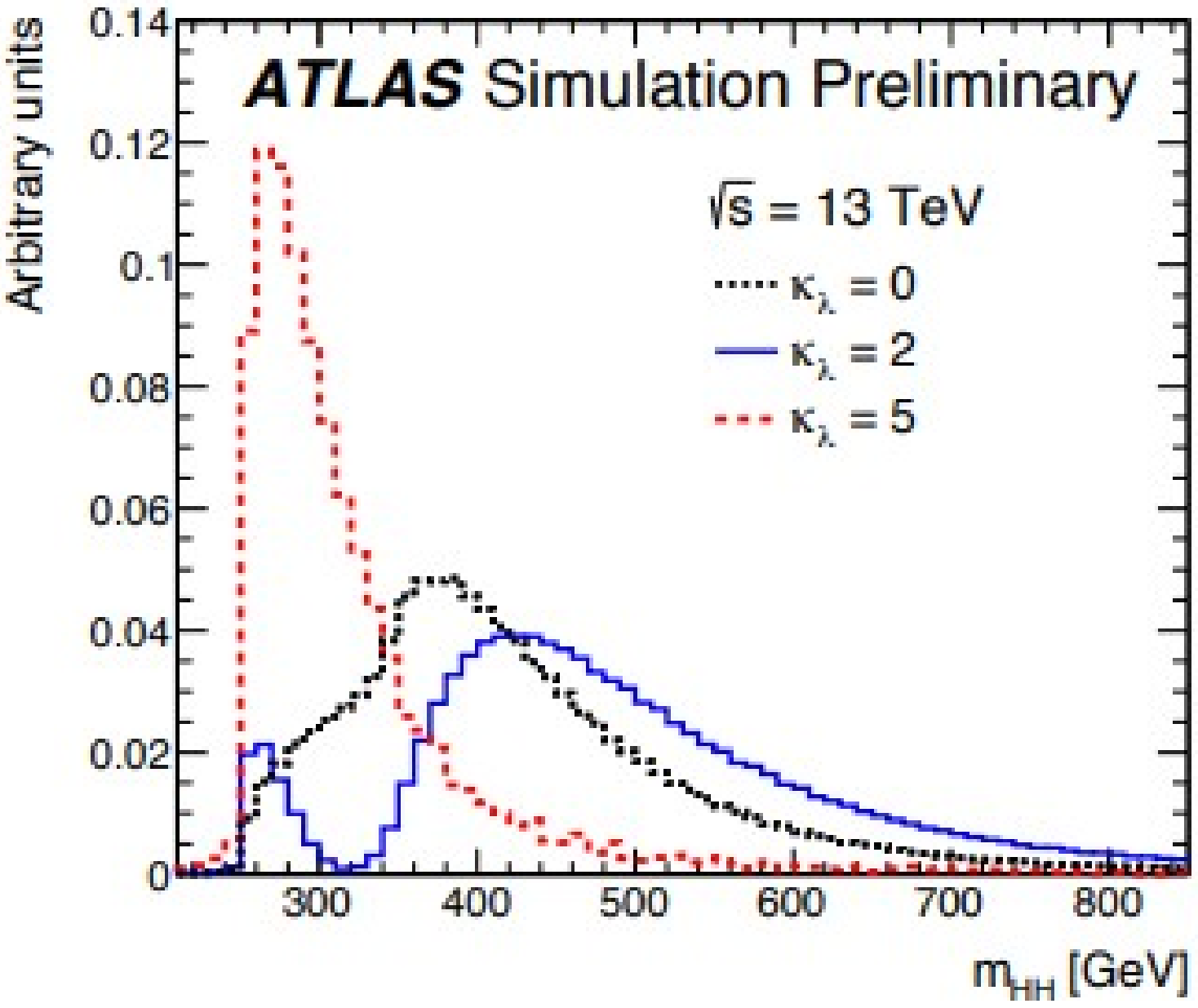

*Figure 3 : Mass distribution of the hh final state for various values of the parameter $\kappa_\lambda$ .*

For ggF→H(320)->HH→4b the situation is much more favorable. One expects σggFHH~300 fb. Adding A(420)→H(320)Z and including the BR into 4b (see appendix) this gives 140 fb. With 1% eff, one expects 170 events over a background of ~20000 events, which should result in a 1.2 s.d. effect, barely visible in figure 2. Figure 3 suggests that an excess around 300 GeV due to H(320)→HH could lead to a misinterpretation as due to a $\kappa_\lambda$ anomaly.

This excess could be reinforced and made fully significant by adding a mass selection of A(151)A(151) which would keep a total number of 700 events, implying a ~5 s.d. excess.

Given the huge QCD background involved in this analysis, it is essential to estimate this background with a reliable procedure, preferably using **data based methods.** This issue has been addressed by LHC collaborations and seems to be under control [7].

A better approach to observe this signal is to search for A(420)→H(320)+Z. We already know [2] that a 3.8 s.d. excess has been reached without including A(151)A(151). Even taking into account an increased background one therefore expects an excess **above ~10 s.d** by adding H(320)->A(150)A(150)**.**

# III. Cross sections in e+e-

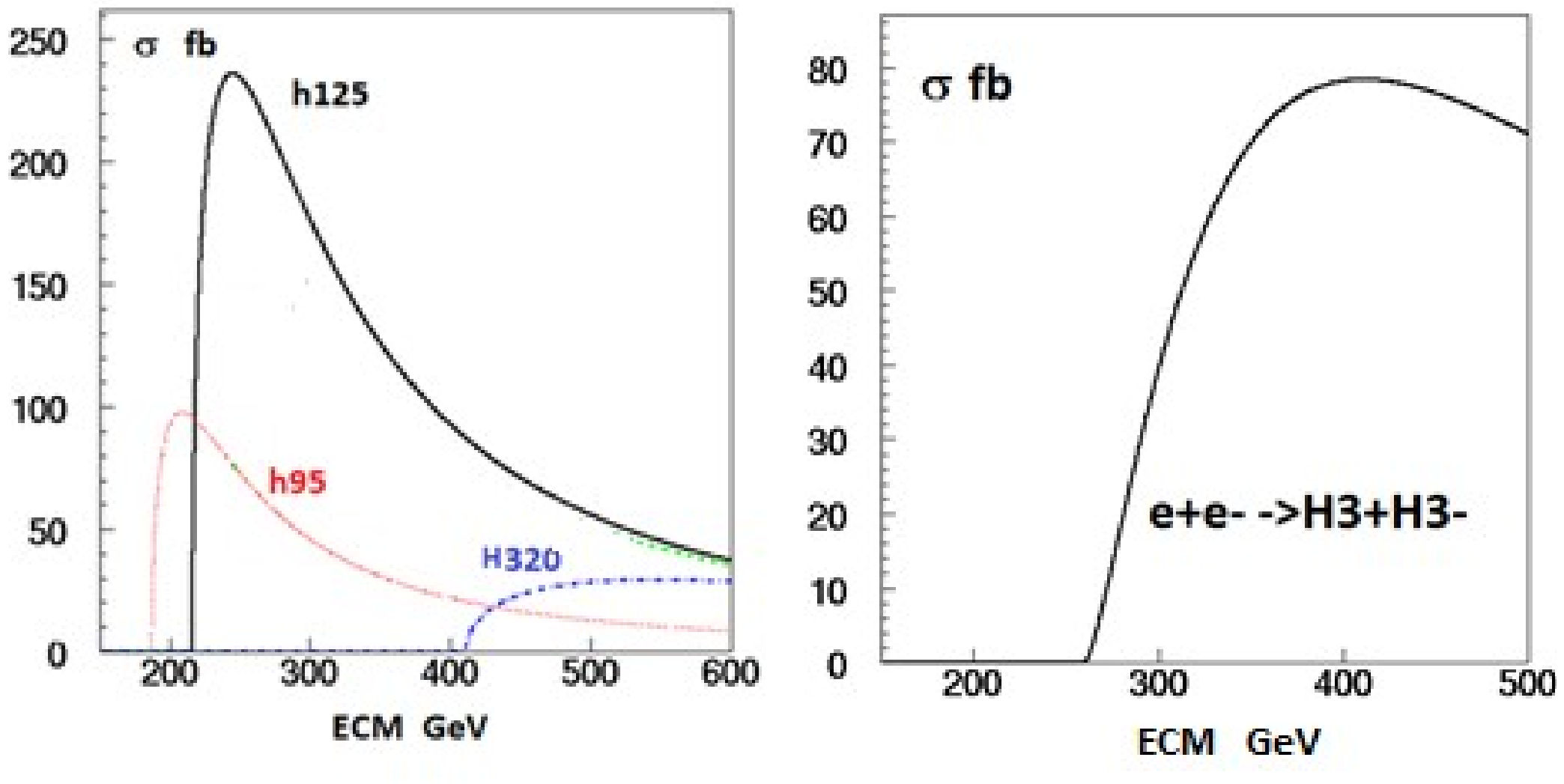

*Figure 4: Predicted cross sections for the lightest scalars.*

These curves are from [1]. The coupling of H(320) to ZZ is sufficiently large that the process e+e- → H(320)Z has a substantial cross section as indicated in figure 4, where the lightest GM states are reported. An ILC operating at 500 GeV will collect ~4000 fb-1, producing more than $10^5$ such events.

An e+e- machine should be able to separate the contributions of A(151)A(151) and HH in bbbb, benefiting from kinematical constraints. It would presumably also isolate the h(95)h(95) channel.

## IV A road to discoveries

What to do next ? It seems that the best strategy would be to update the A→ZH(320)->bbbbZ analysis by including a mass selection for the A(151)A(151) contribution. One would kill 3 birds with the same bullet since the analysis would:

1. Confirm H(320)
2. Confirm A(420)
3. Confirm A(151)

One could also try discovering A(420)→H+(130)H-(130)Z→bcbcZ, recalling that 36% of H(320) decays go into H+H-. The road to di-Higgs inclusive discovery into bbbb is also opened but perhaps more arduous.

## IV. Summary and conclusions

This brief analysis shows that the bbbb topology is likely to provide a proof of the presence of **triple Higgs couplings** like H(320)→A(151)A(151) and HH→bbbb, which would constitute the strongest evidence for BSM physics at LHC, a first step in exploring the properties of the Higgs potential within the GM model. An other access to such couplings would be the direct identification of H++(450)→H+(130)H+(130) which is dominant according to our analysis [1].

We have identified two potential sources providing H(320):

- The cascade A(420)→H(320)Z which gave the first evidence for H(320), with a cross section of about 700 fb
- The gluon fusion process ggF→H(320) with a cross section of 2000 fb

These two channels need therefore to be analysed with the assumption that the bbbb final state can also originate from **A(151)A(151).**

For what concerns the channel **A(420)→H(320)Z**, adding HH and A(151)A(151) a very convincing proof of the resonances A(420), H(320) and A(151) should be reached in the **bbbbZ final state.**

**We therefore predict that by adding HH and A(151)A(151) contributions, the di-Higgs search should deliver a fully significant signal for H(320), well above 5 s.d., which would constitute the most convincing signal for proving BSM physics at LHC.**

Note that these results do not require the HL-LHC phase but can be reached with the RUN2 data.

It is however fair to add a word of caution since this prediction is based on results coming from several indications which may vanish and from a chain of reasoning which is quite involved. Recall, in particular that our evaluation of ggF→H(320) relies on an indirect evaluation of the top Yukawa coupling of H(320) through the 'matrix method' developed in [1].

We also rely on an indirect derivation of the triple scalar coupling H(320)→A(151)A(151) which follows from an estimate of the BR of H++(450) into W+W+ provided by the unitarity sum rules.

It is however fair to say that these predictions rely on a large set of **direct observations** provided by LHC data which are consistently described within a **solid phenomenological framework** which should guide future searches.

This framework may however still require further extensions, as suggested by a tension between the value of the **vacuum expectation u** deduced from the SR and the indirect determination of this quantity from the B physics constraints. The SUSY version of GM, **SGM**, described in section V of the appendix, is a potential candidate to remedy to this tension. It contains a larger number of scalars to be discovered with HL-LHC.

### Acknowledgments

*This work relies on the phenomenological inputs provided by our colleagues Anirban Kundu, Gilbert Moultaka and Poulami Mondal who are warmly thanked for their many contributions.*
*F.R. is grateful to Andreas Crivellin for numerous and very useful email exchanges during the Moriond 2024 meeting.*
*Thanks to Mioaran Lu from ATLAS for following closely this work.*
*We are grateful to Ulrich Ellwanger for providing very useful informations, in particular on the decays of neutralinos in NMSSM.*

# Appendix

## I. Inputs for H(320)

Figure 5 from [2] shows the evidence for the cascade A(420)→H(320)Z→bbbbℓ+ℓ-. The local significance reaches 3.8 s.d. Recall [1] that A(420) is also observed in top pairs and through the transition H(650)→A(420)Z→ttℓ+ℓ-. Based on an MSSM interpretation which predicts mA>mH, this analysis assumes, without proof, A(650)→H(420)Z, at variance with our interpretation within the GM model.

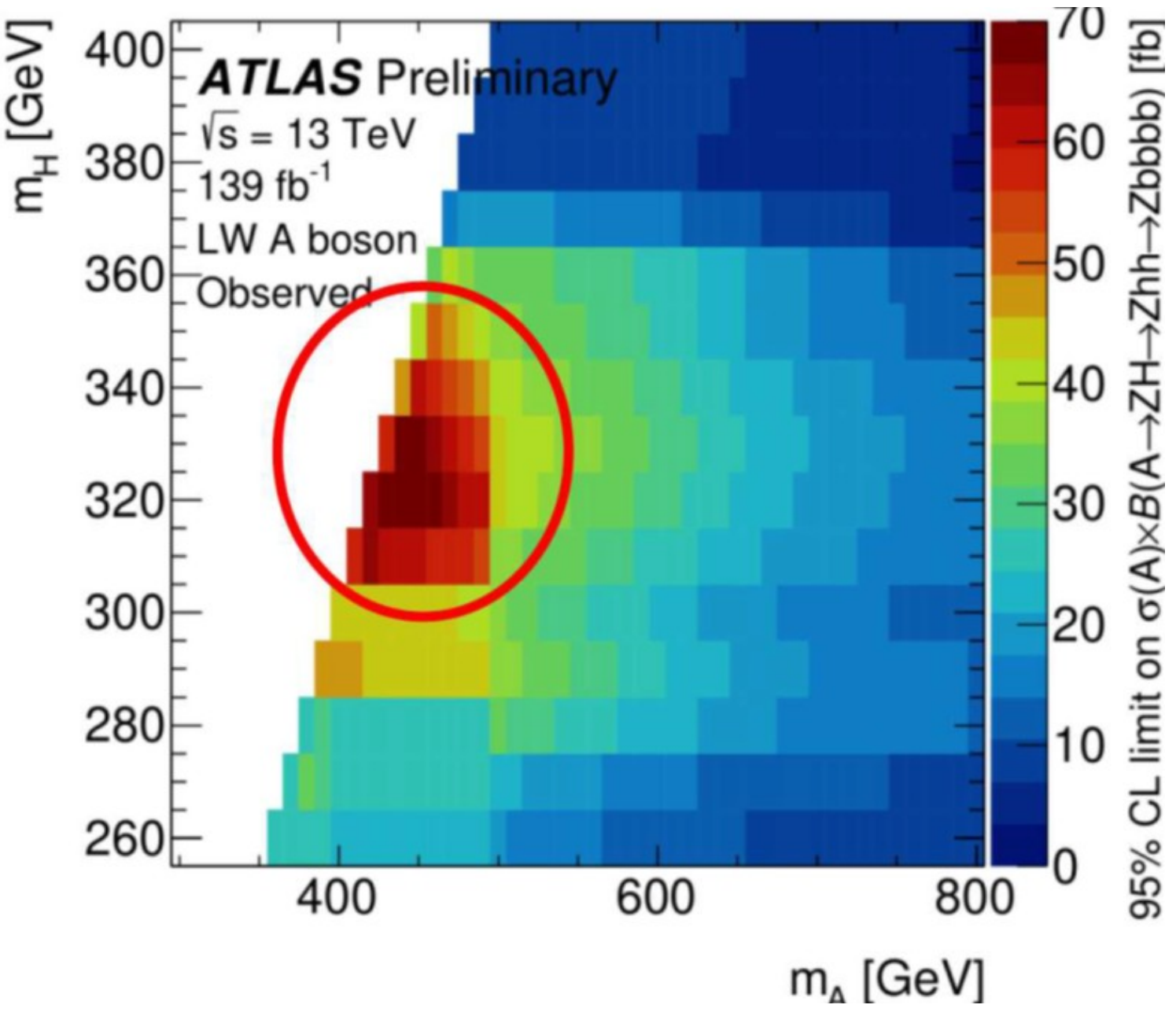

*Figure 5: Observed upper bounds at 95% CL on s(A)xBR(A→ZH(320)→ZHH→Zbbbb in the (mA,mH) plane for a large width A(420) boson.*

If one assumes that H(320) is identified to H5 from the GM fivetuple of GM, it decays into A(151)A(151) and H+(130)H-(130) and it decouples from singlet states. Mixing deeply modifies this picture, as will shall see in section 3, allowing decays into HH and, to a lesser degree, into h(95)h(95). One predicts **AA/H+H-/HH/hh=45%/36%/17.5%/0.5%** with **Γtot=220 GeV.**

It is likely that the mass and the width deduced from from A(420)→HZ, are underestimated given the constraints of this decay which is close to the kinematical limit.

Assuming H(320)→HH, [2] sets a limit **σ(A)BR(A->H(320)Z→bbbb)<70 fb with 95% CL**. The excess above background suggests that this cross section is ~ 35 fb.Taking into account the BR into hh and the BR of hh into 4b, this amounts to a cross section σA->H(320)Z~700 fb, to be compared to σggF->H(320)=2000 fb as predicted by the matrix approach (see section III)

If A(420) couples maximally to the top quark, one expects σggF->A(420)~40 pb and a width into top pairs ~10 GeV. This means that **BR(A420→HZ)=2%**. This small value does not come as a surprise given that the reaction A(420)→H(320)Z is close to the kinematical limit. One may speculate that **A(420)->HZ** is likely to be observable given the more favorable phase space. There were indeed indication for such a reaction [15].

## II. Indications for A(151) ⟶ γγ+tags

[3] has collected the various indications for a resonance in two photons which can be observed using various tagging signatures as shown in figure 6. The combination of these observations reaches **4.8 s.d.**

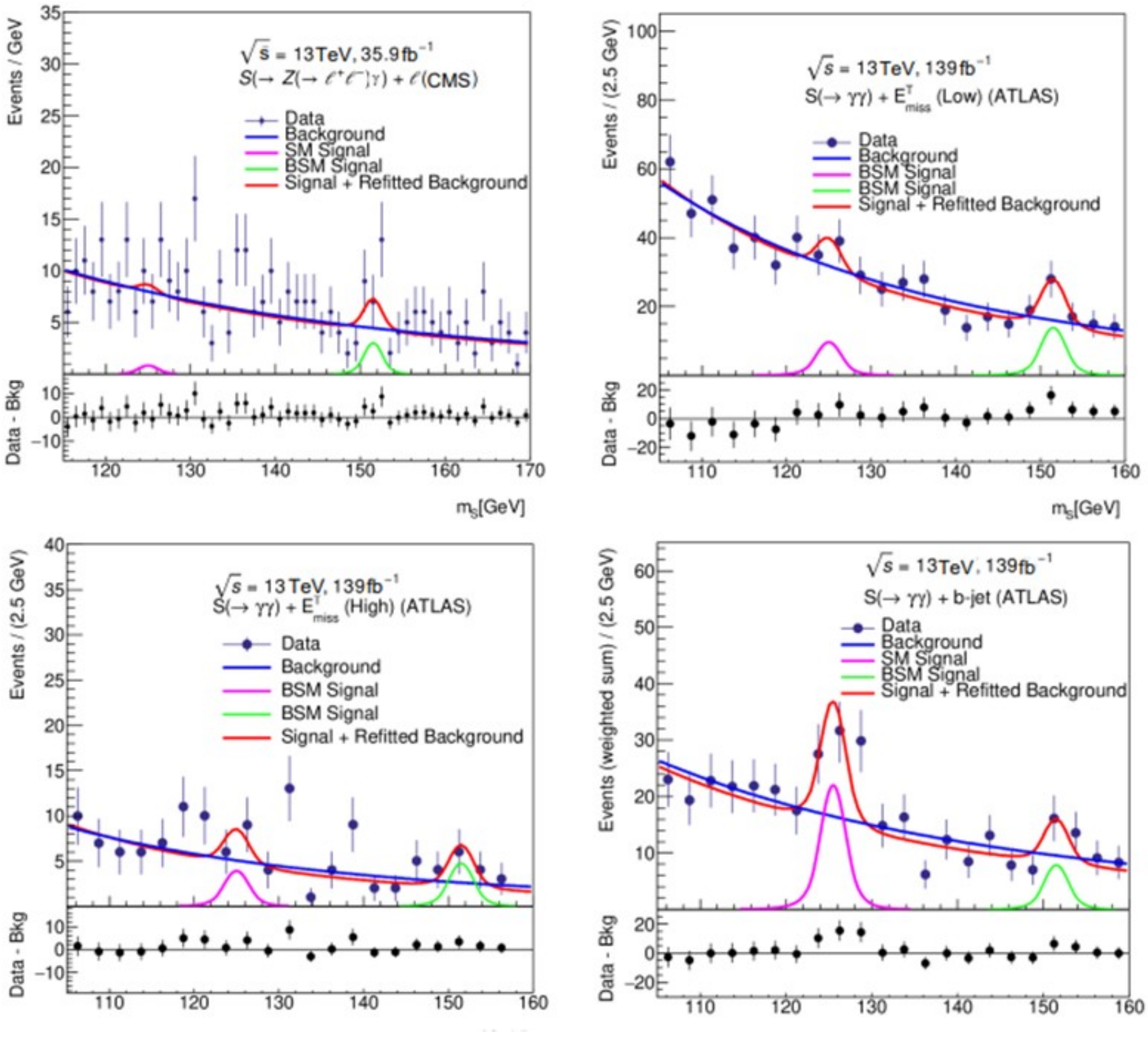

*Figure 6: γγ mass distributions observed by ATLAS in CMS when adding a tagging requirement.*

These observations appear compatible with the processes:

**H(320)(Z)→A(151)A(151)→ γγ +ττ or bb (bb,νν,ℓ+ℓ-)**

A recent indication [7] for the ττ interpretation is shown in figure 7.

A back of the envelope estimate including the following table derived from GM, allows to quantitatively comfort our interpretation of the origin of these tagged events.

| Channel | bb | ττ | cc | gg | γγ |
|---|---|---|---|---|---|
| BR(A151) % | 74 | 8 | 3.7 | 15 | 0.13 |

Note the difference with respect to H(125) where the BR in bb is below 60% given the ZZ/WW contributions.

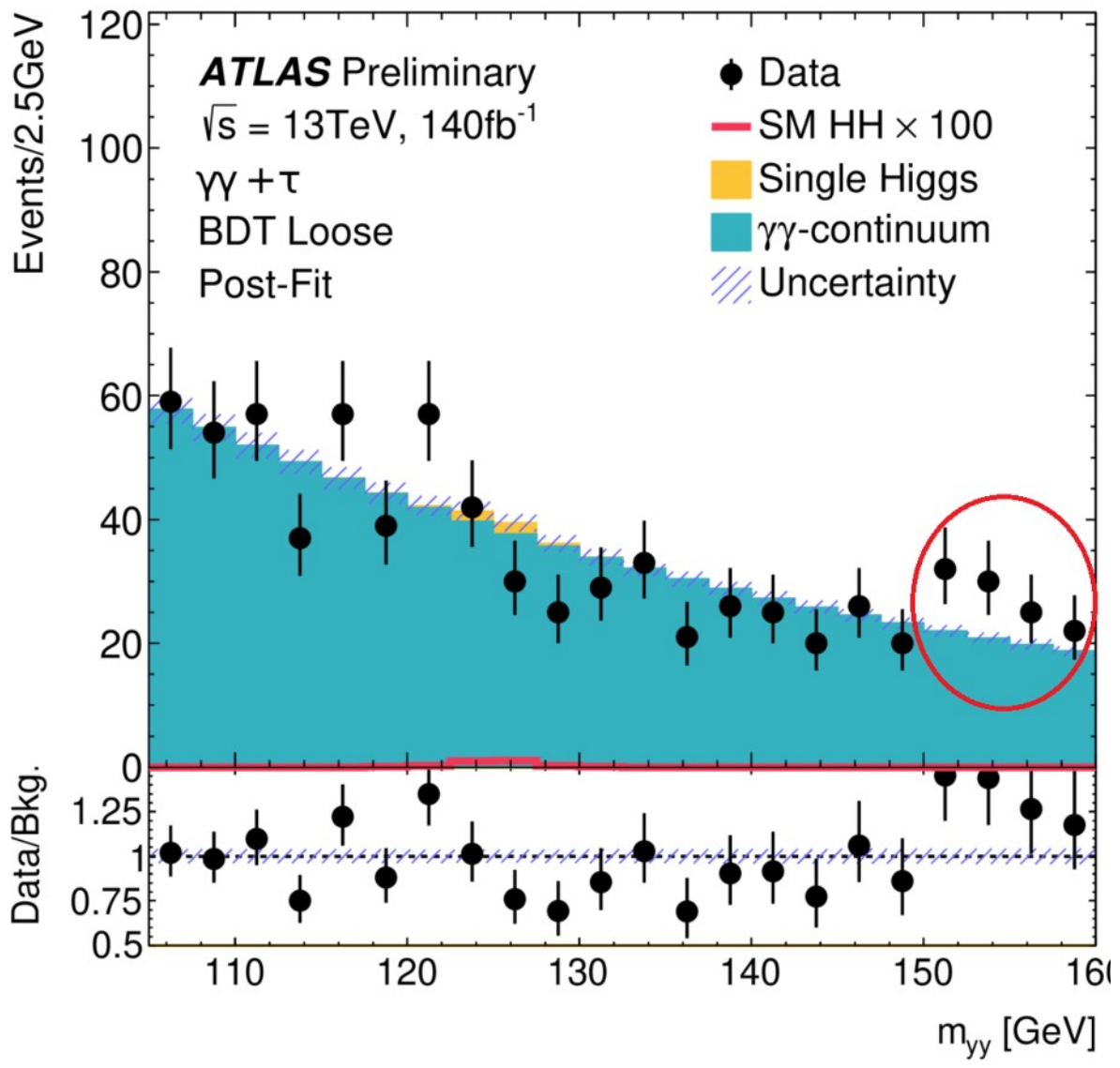

*Figure 7: $\gamma\gamma$ mass distribution observed in [7] when requiring the presence of a $\tau$ particle.*

Since A(151) has not been observed in the ZZ mode, it seems reasonable to assume that it is a **CP odd scalar**. This fits perfectly to the GM description where H+(130) and A(151) pertain to the same triplet.

It was suggested [3] that this resonance is also observed in the WW mode. However one should not forget that the mass resolution for this mode, observed in leptonic modes with neutrinos, is poor and that there is no clear mass separation with the SM mode. Therefore this interpretation remains an open question.

## III. A global interpretation of LHC indications for scalars

This section intends to collect and summarize inputs from [1] which are needed for the present analysis.

The diagram below summarizes our present list of significant LHC findings [1]. This reference also gives the list of papers providing the various indications and their significance with a marked emphasis on H(650) which is the driving input for the sum Haber et al. rules.

The table below summarizes a global interpretation of a plethora of indications observed at LHC in terms of **an extended Georgi Machacek model.** One is able to identify all states predicted by the original GM model.

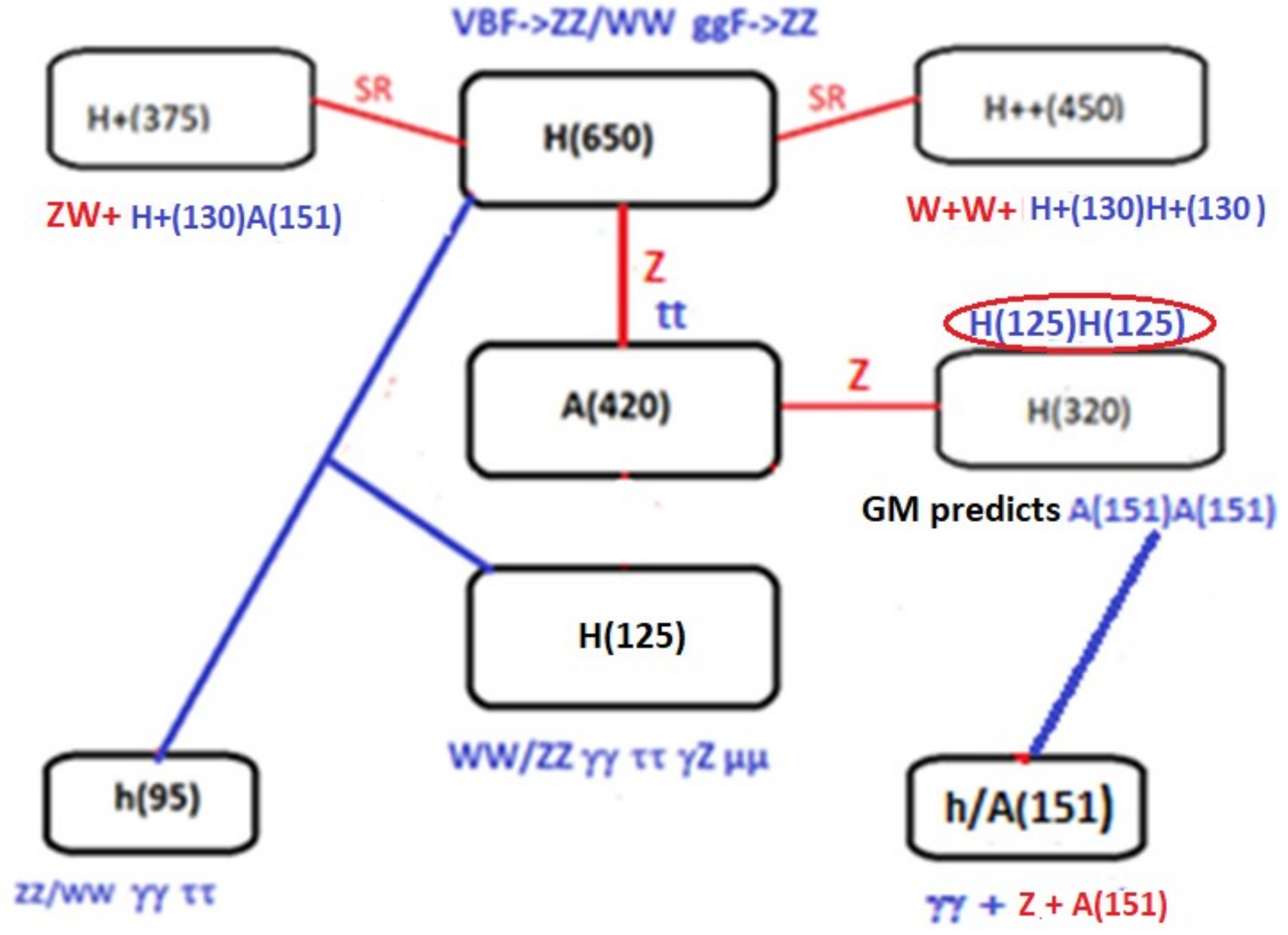

To interpret H(650) properties, one has to add an extra doublet to this model, predicting 3 additional states, two of them fitting nicely in this extension, while there is no clear evidence for the predicted extra charged scalar **H+(x)**. We further discuss this issue in section IV.

| | | | | |
|---|---|---|---|---|
| **GM states** | **Isosinglet** | **h95** | **h125** | |
| | **Isotriplet** | **a151** | **h+130** | |
| | **Isofiveplet** | **H320** | **H+375** | **H++450** |
| **Additional** | **Extradoublet** | **A420** | **H650** | **H+ ?** |

One has four CP even scalars and the following table, derived as explained in [1], indicates that the physical states are mixtures of the doublets and triplet fields, including the SM H(125).

| | 1 | 2 | 3 | 4 | Ytt/SM | ZZ/SM | WW/SM |
|---|---|---|---|---|---|---|---|
| | $\phi1$ | $\phi2$ | $\chi$ | $\xi$ | | | |
| h95 | 0.10 | - 0.56 | 0 | 0.80 | - 0.96 | -0.34 | 0.60 |
| H125 | 0.58 | 0.58 | 0.47 | 0.33 | ~1 | ~1 | ~1 |
| H320 | 0.27 | 0.34 | -0.80 | 0.35 | 0.50 | -1.20 | -0.40 |
| H650 | 0.79 | -0.50 | -0.1 | -0.34 | -0.90 | -0.60 | -0.90 |

$v_1$=-30 GeV, $v_2$=102 GeV are vacuum expectations for the for the two doublets, u=70 GeV for the triplets. One has v=174 GeV=sqrt($v_1^2+v_2^2+4u^2$). With a **type I solution** assumed for the Yukawa couplings, one has *Yi/SM=xi2mt/v*. For the HiWW coupling normalised to the SM value, one has:

$$Hi_{WW}/SM=(x_{i1}v_1+x_{i2}v_2)/v+(2x_{i3}+2\sqrt{2}x_{i4})u/v$$

Note that we have assumed that the **couplings of H125 are all positive** as is the case for the SM. LHC data [8] only set a constraint on the relative signs of HWW and HZZ couplings which implies that both could be negative. This has implications on the predictions of the two photon decays of h95 and H125.

Missing quantities concerning H(320) and h(95) (see the colored squares above) can be predicted, in particular for what concerns H(320).

The decay widths of h95 and H125 into two photons normalized to the SM are predicted by this matrix close to 1 taking into account only WW and tt contributions, to be compared to **μh125γγ**=1.1**±0.09** and **μh95γγ =0.24±0.09** [12] measured by LHC. Adding the charged scalar contributions requires a full knowledge of the scalar potential in the extended GM which is not yet available. One can predict that there should be a cancellation of these terms for H125 and a substantial negative contribution for h95.

**III.1 The H1-H5 mixing issue**

The matrix method gives critical informations about the four neutral candidates. Generally speaking their properties **markedly differ from the isospin proper states of GM** which has very important experimental consequences. It suggests that **custodial symmetry** only preserves part of the Lagrangian needed to protect the ρ parameter, but is largely violated elsewhere [9]. As an example, if H(320) would strictly behave as a proper state of H5, it would not couple to H(125)H(125) and H(320) would not have been discovered in the HH→bbbb mode. Mass degeneracy within isomultiplets is rather well satisfied except in the A(420)/H(650) case. As pointed out in [9] there are ways to extend GM to lift mass degeneracy still preserving the ρ parameter constraint.

H650 is dominated by the two doublet components, and therefore is not a likely candidate for the 5-plet. H320, which is dominated by a χ component, naturally fits into a 5-plet, with a mass compatible with H+(375) and H++(450).

The SM H reads **H(125)=0.58(φ1+φ2)+0.58H'1,** where the triplet fields appear in the singlet combination **H'1=√1/3ξ+√2/3χ** which explains why they have not induced significant deviations from the SM. These deviations could, however, show up at the loop level in two photon final states.

One has **h95 =-0.56φ2+0.5H'1+0.7H5.** This is probably the **most striking result** of the matrix method which requires a detailed interpretation.

Recall that GM is unable to predict the Higgs mass nor the value of the ρ parameter since both quantities receive divergent loop corrections. This feature causes worries about the **naturalness problem** as nicely discussed in [10]. Insuring mH=125 GeV, that is well below 1 TeV, carries the same degree of fine tuning of the theory as protecting |ρ-1| to be <1 %. In reality, this deviation is far below this bound – ρ=1.00037±0.00023 – suggesting that this F.T. issue is much more severe than the mH one.

Understanding the large α15 mixing which is manifest in the case of h95, offers similar challenges since this mixing also receives divergent loop contributions.

These three issues can be formally solved by **standard renormalization procedures** but the finite residual quantity cannot be computed within GM alone but require the knowledge of the **u.v. completion** of this phenomenology. For the time being one can only state that **mH**, ρ and α15 are **measurable experimental quantities**. With the matrix method we were able to measure the third one but we cannot interpret quantitatively this mixing as being too large or too small with respect to expectations.

One may hope (see section V) that SUSY coud eliminate theses divergences and predict these parameters, as was the case for mH, but this remains to be proven. Many other questions come to mind, like the following one: how is is that H1-H5 mixing does not affect H125 ? Note in passing that H1-H3 is forbidden by CP conservation since H3 contains a CP-odd scalar [10].

### III.2 The unitarity sum rules

In the **charged sector**, the **unitarity sum rules from** [11] relate W+W- and ZZ couplings of the neutral scalars to those of the **charged scalars**. One derives: $g^2_{H++WW}/SM=1.3\pm0.4$ $g^2_{H+ZW}/SM=0.8\pm0.2$ hence the partial width of elastic modes: $\Gamma_{H++WW(450)}$=15±5 GeV $\Gamma_{H+ZW(375)}$=12±4 GeV.

From these and the SR, one can deduce the total VBF cross section, the elastic BR and the total widths as given in the following table:

| Channel | $\sigma_{VBF}$ fb | $\sigma_{VBF}$ VV fb | BR(VV) % | $\Gamma$tot GeV |
|---|---|---|---|---|
| H++(450) | 830 | 75 | 9±4 | 160 |
| H+(375) | 810 | 125 | 15±8 | 80 |

**These are model independent results** derived from measurements [1]. BR into W+W+ and ZW+ are below 20%, implying dominant contributions of HV and HH final states. Indications for H+(130)->bc and for A(151)->$\gamma\gamma$ as GM triplets comfort this interpretation.

Within GM, one can deduce the vacuum expectation of the two triplets constituents, called hereafter **u,** and the related quantity $\mathbf{s_H}$. This allows to estimate BR(VH) which turns out to be of the same order as BR(VV), suggesting the need for a large BR(HH), as summarized in the following table.

| Channel | u GeV | $s_H$ | BR(VV) % | BR(VH) % |
|---|---|---|---|---|
| H++ | 70±12 | 0.80±0.1 | 9 | 12.5 |
| H+ | 80±13 | 0.90±0.2 | 15 | 17 |

One can then adjust the GM parameters such that the HH contributions complete the missing BR. Together with the observed properties of H(125) and h(95), this allows an extraction of the other parameters of this model [1], a way to **fully reconstruct the Higgs potential** in the GM model. This results in the following GM parameters[3]:

| u GeV | m5 GeV | $\lambda$1 | $\lambda$2 | $\lambda$3 | $\lambda$4 | $\lambda$5 | M1 GeV | M2 GeV |
|---|---|---|---|---|---|---|---|---|
| 70 | 400 | 0.07 | -1.4 | -1.06 | 1.25 | -6.3 | 950 | 400 |

Note that the large **values of the parameters M1 and M2** originate from the requirement that H++→H+H+ is dominant over W+W+. They naturally induce **large triple Higgs couplings.**

3 Beware that these parameters follow the conventions from [1] while other authors use the mass parameters M1 & M2 with opposite signs and reverse $\lambda$2 and $\lambda$4.

The large value of $\lambda 5$ which causes some concern for perturbative unitarity [24] and may require taking into account HO corrections. Also this limit has to be reconsidered in an extended GM model.

### III.3 The triple Higgs coupling

One can extract the triple H125 coupling, H*→HH, normalized to the SM, $\boldsymbol{\kappa_\lambda}$**=5.** This result is compatible with LHC measurements as shown in figures 8, 9 and 10. It also coincides with the reflection expected from the resonant decay H(320)→H(125)H(125) (figure 3).

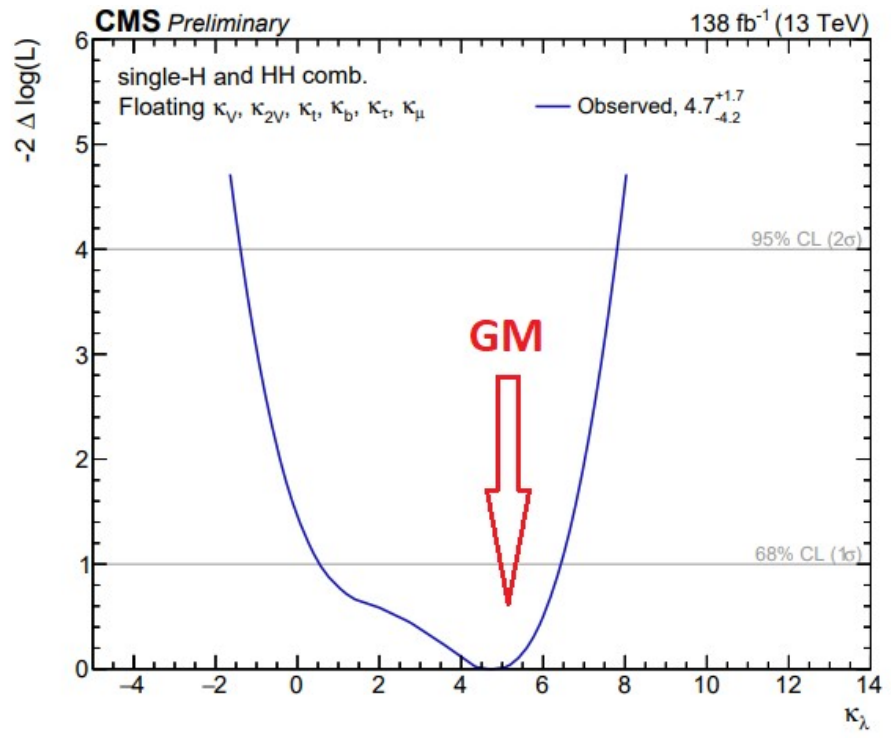

*Figure 8 : A recent compilation from CMS [13] on the self-coupling parameter $\kappa_\lambda$ normalised to the SM. Observed likelihood scans of κλ assuming κV, κ2V, κt , κb , κτ , and κμ as unconstrained nuisance parameters.*

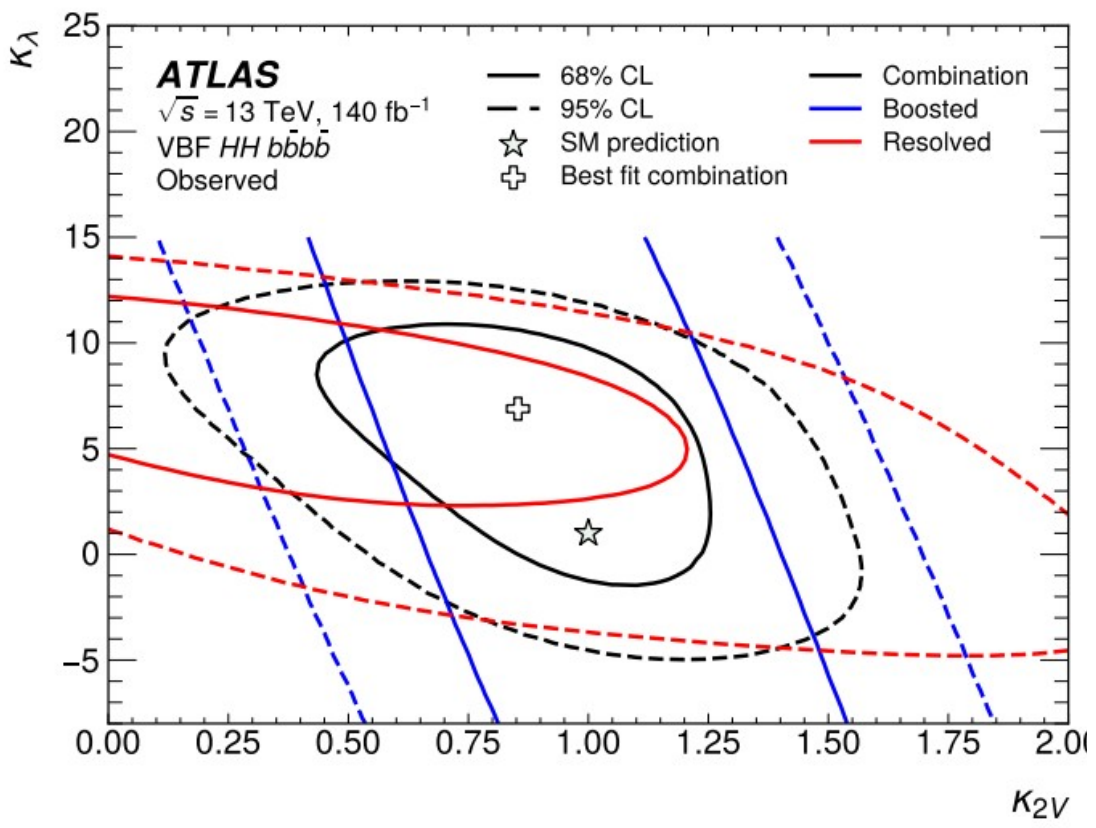

*Figure 9: Observed likelihood contours in the $\kappa_\lambda$ k2V plane. The black ellipse is the result of a combination of resolved and unresolved VBF bbbb analyses.*

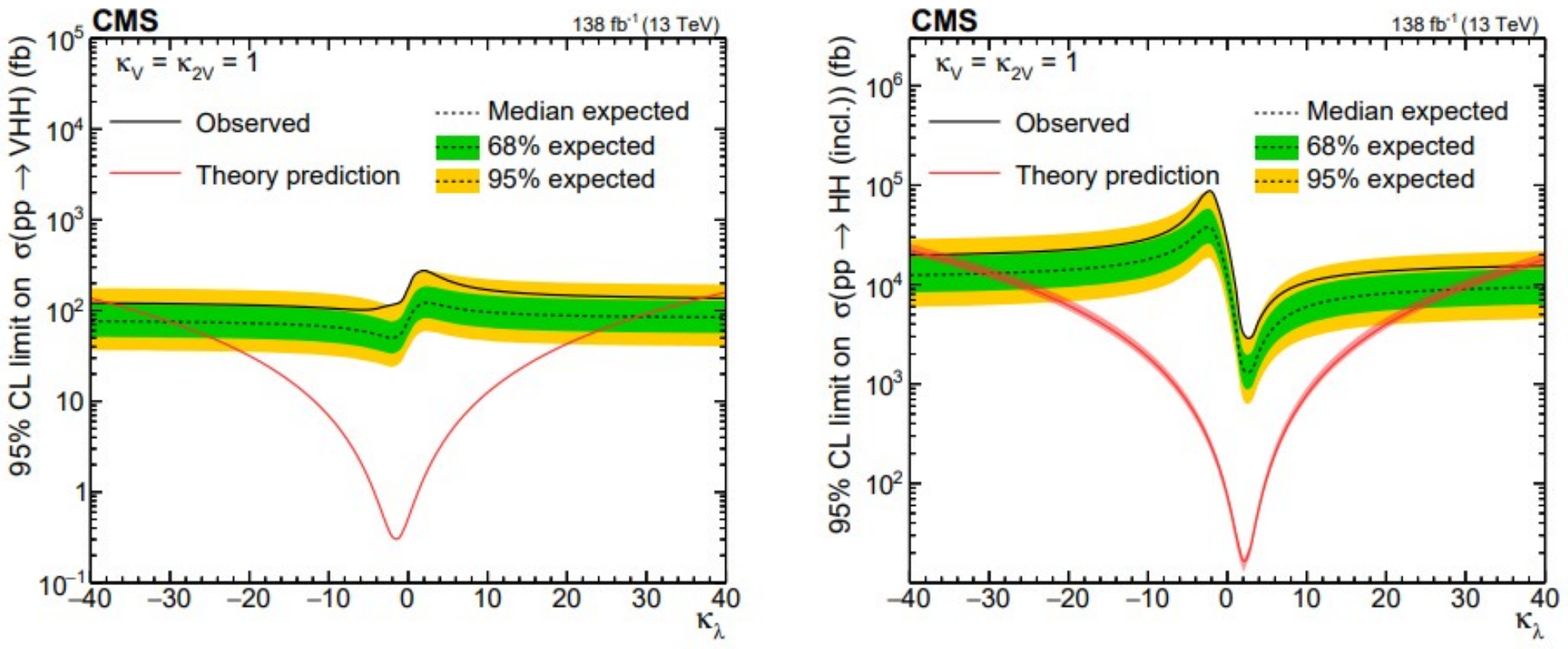

*Figure 10: 95% CL limit on VHH (left) and HH (right) cross sections for SM couplings for κV=κ2V=1.*

# IV The missing H+

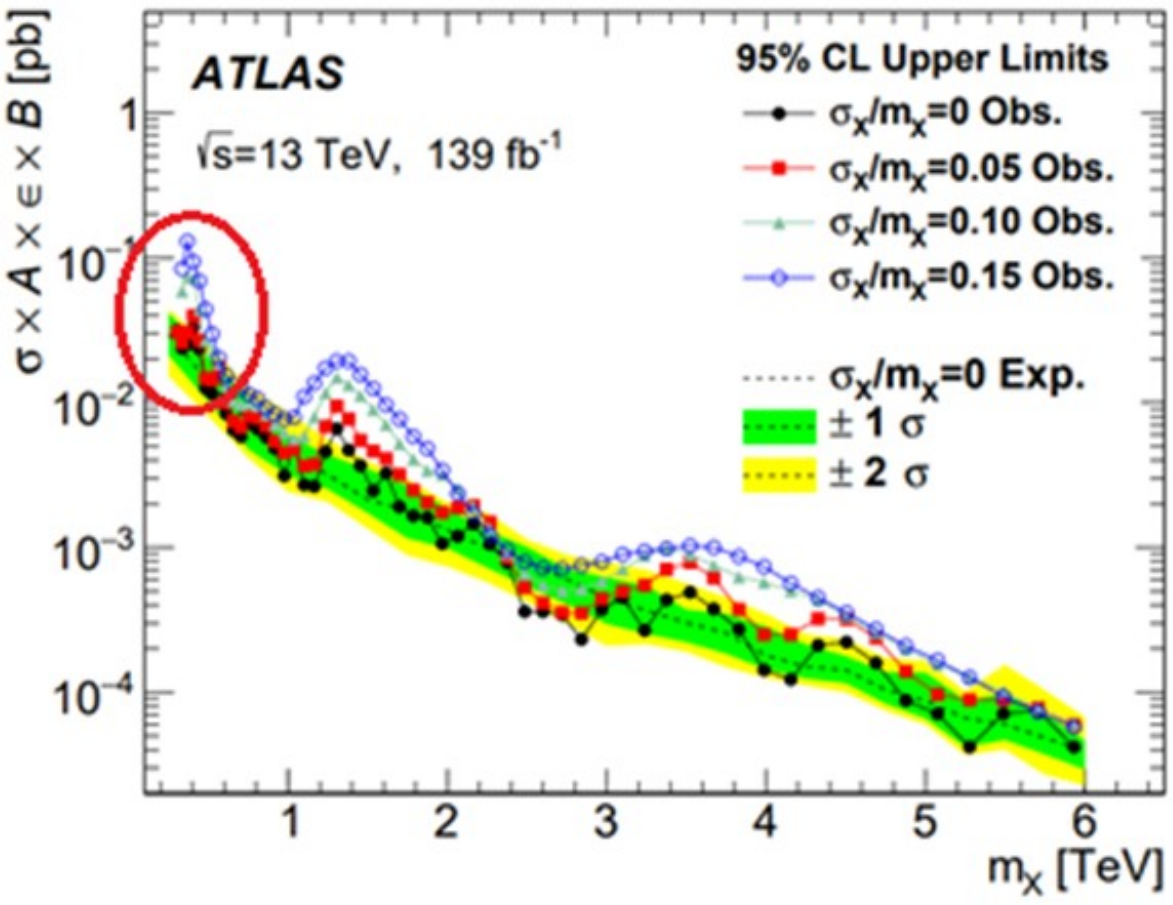

*Figure 11: Jet-Jet mass recoiling to a large pt lepton.*

In [14] there is a candidate for H+→ttW+ provided by an inclusive search for heavy resonances decaying into two jets accompanied by a high pt lepton interpreted as coming from a W. The reconstructed mass of the two jets is compatible with A(420)→tt and therefore is a candidate for **H+→A(420)W+.**

Figure 11 shows the largest effect for large values of $\sigma_X/m_X$ implying that A(420) has a total width of order 100 GeV. This width could also be due to a poor mass resolution.

This process mirrors the process observed by [16] with our interpretation:

**H(650)→A(420)Z→ttZ**

It is likely that these two particles have about the same mass. *T*his finding also provides an interpretation for the excess reported for the ttW+ and ttW- channels [15] shown in figure 12.

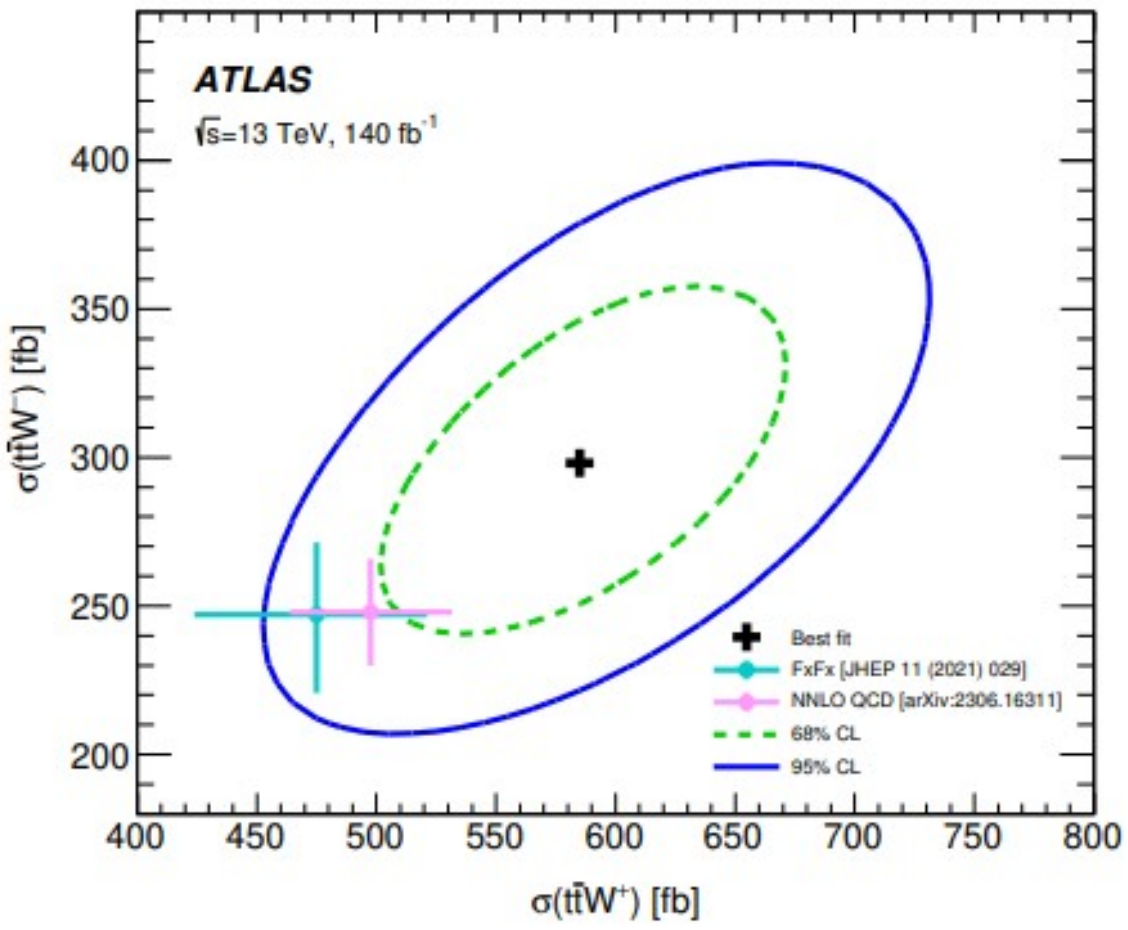

*Figure 12: Measurements of ttW+ and ttW- cross sections [15]*

Within GM, a plausible production mechanism is the fusion W+Z→H+. This would explain that the excess is more pronounced for ttW+, given that energetic W are emitted by the valence quarks of the incoming proton:   p→u→W+ / p→d→W- = 2

One should therefore also observe H+→ZW+ provided that the BR is not too small.

This interpretation has a practical consequence on the SR: this extra H+ should contribute and reduce accordingly the value of the vacuum expectation u, which would be welcome given the constraints due to B physics (see figure 13 of next section).

The contribution from H(650)→AZ->ttZ shows no impact on σ(ttZ) measurements. This difference, still marginally significant, indicates that σ(H650→AZ)BR(ttZ) is smaller than σ(H+→AW+)BR(ttW+) which, to be understood, requires a precise determination of the composition of H+ not yet achieved so far.

# V SUSY extension of GM

Accommodating H(650) has required adding an extra doublet. This move recalls what happens when going from **SM to MSSM** under the influence of SUSY.

Is this sufficient to accommodate present observables? This does not seem to be the case judging from the B physics constraints shown in figure 13. The value of u~70-80 GeV derived from the SR seems too large with respect to the upper bounds derived from B physics measurements. It is however fair to say that this evaluation is based on GM alone and does not include e-GM nor GMS.

In the WW->ZZ SR we have not taken into account the contribution from the heavy H+ predicted by e-GM. This particle seems to couple to ZW+ and therefore should also contribute to this SR, therefore reducing the value of u. A SUSY version of GM, SGM [20,21], would predict an extra H++, which also reduces the value of u in the WW SR, re-establishing the consistence between the results of the two values. As recalled in [20], GM loop contributions give divergent corrections to the ρ parameter and the hope, not yet materialized, is that these divergences could be eliminated within GMS, as is the case for loop corrections to the Higgs mass.

It is therefore tempting to predict that SGM could be at work, doubling the number of scalars to be discovered at LHC with respect to GM, not to speak of the SUSY sector itself.

If so, one expects the following states :

- Three singlets h h' a
- Two doublets H+ H A
- 1 Triplet H+ A
- 1 Fiveplet scalar H++ H+ H
- 1 Fiveplet pseudo H'++ H'+ A

**In total 20 states : 2H++ 2H- - 4 H+ 4H- 4A 4H**

Therefore no change in the neutral CP even sector within e-GM, but with u likely to be reduced by SR contributions due to additional charged scalars.

Note that for the WW->WW SR, one may think that only the doublet and scalar fiveplet isospin states contribute but recall that the physical states substantially differ from the isospin states, therefore the four neutral CP even physical states will contribute as well as the two doubly charged physical states.

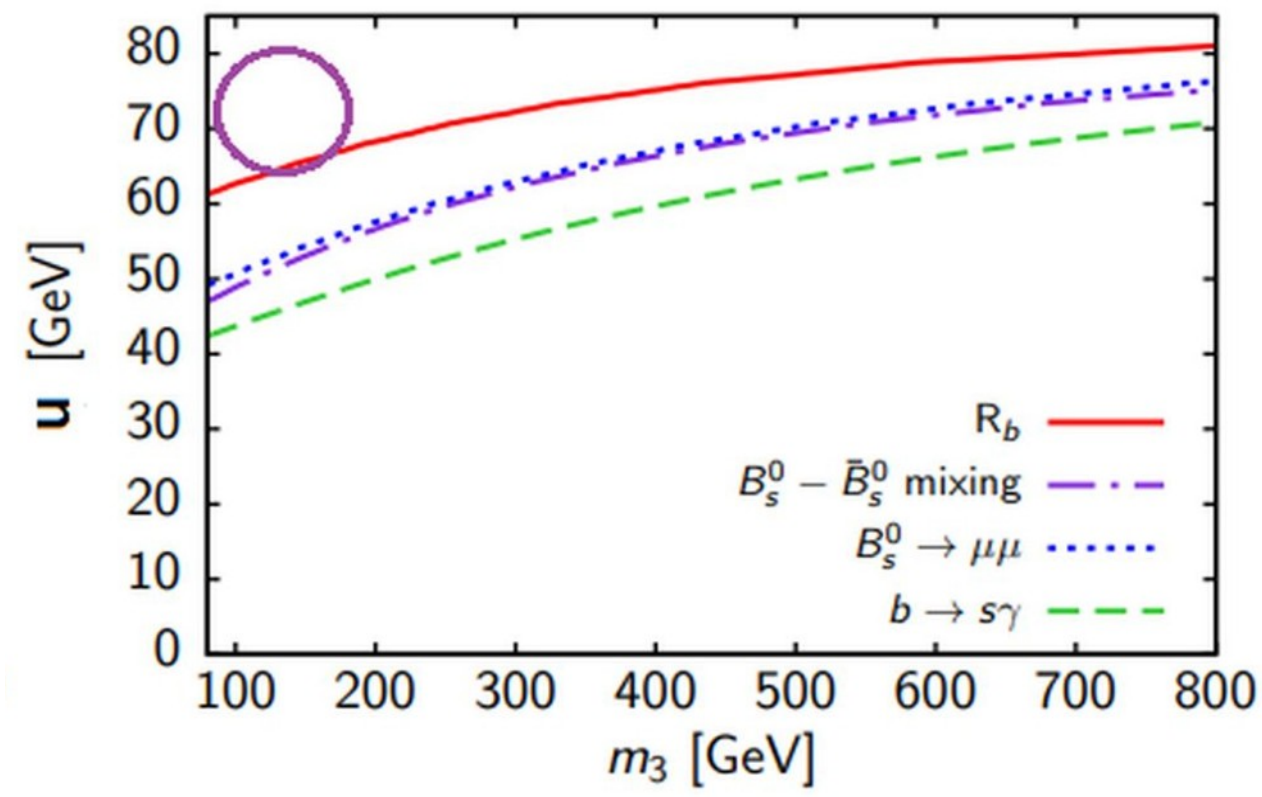

*Figure 13: This picture, taken from [20], indicates the upper limits (2 s.d. level) derived on the parameter u of the GM model from B physics PM. The ellipse shows our prediction.*

SGM provides all the "goodies" of SUSY:

- **Perturbativity, computability**
- **GUT unification of forces**
- **Insensitivity to quadratic divergences** (Higgs mass, $\rho$ parameter,H1-H5 mixing...)
- **EWSB** naturally generated
- **DM** singlino candidate which, as for NMSSM [23], escapes to direct detection experiments
- **MH=125 GeV** accommodated with less "tension" on **stop masses [20]** (6 TeV in MSSM)

To summarize, there are indications that e-GM is incomplete and that the GM spectrum could be even richer than suggested by the various indications presently identified in LHC data. SGM predicts that there c could be **two additional CP odd scalars, one H+ and one H++, to be discovered** and of course a renewed motivation to search for SUSY particles. The next obvious question is: **where is SUSY** ? As well-known, there are scenarios where SUSY has so far escaped to LHC searches due to mass degeneracies, the so called 'compressed-spectrum' case, which explains an absence of signal [23] at LHC. This reference uses an NMSSM interpretation of SUSY which, as in SGM, contains singlets. It recalls that there are weak but coincidental indications from ATLAS and CMS for quasi-degenerate gauginos. This model predicts three neutralinos and one chargino clustering around 250 GeV, a scenario eminently favourable to an ILC collider reaching at least 500 GeV.

**NMSSM** and **SGM** predict a **Higgs singlet** and [23] claims that the 95 GeV candidate naturally fits into the NMSSM scheme.

Reconciling these apparently independent descriptions is an **urgent task for phenomenology.**

Firmly proving the presence of SUSY appears very difficult at LHC. Note that in NMSSM the chargino particle of rank 1 decays into a neutralino of rank 2, the later decaying into the singlino plus a Z* or a photon[4]. Would such a decay be traceable in LHC detectors is an open question which of course depends on the mass difference between these two lightest neutralinos. This feature could perhaps provide the **'smoking gun' effect** needed to confirm this process.

Explicitly, one expects DY processes of the type:

$pp \to u\bar{d} \to W^{+*} \to \chi^+\chi_3 \to W^{+*}\chi_2 Z^*\chi_2 \to \ell^+\nu\ell^+\ell^-\gamma\gamma\chi_1\chi_1$ , $\to \ell^+\nu\nu\nu\gamma\gamma\chi_1\chi_1$ , $\to \ell^+\nu q\bar{q}\gamma\gamma\chi_1\chi_1$

$pp \to u\bar{u} \to \gamma^*/Z^* \to \chi^+\chi^- \to W^{+*}\chi_2 W^{-*}\chi_2 \to \ell^+\nu\ell^-\nu\gamma\gamma\chi_1\chi_1$ , $\to \ell^{+/-}\nu q\bar{q}\gamma\gamma\chi_1\chi_1$

4 According to U. Ellwanger [23], in NMSSM $\chi_2$ decays in 10-30 % of the cases into $\gamma\chi_1$, **the rest goes into $Z^*\chi_1$ .**

**$q\bar{q}$**, corresponding to the presence of a **mono-jet**, has the highest BR but with large background. This background could be significantly reduced by requiring one or two accompanying photons or monojets.

Cross sections are large, at the >200 fb level, as shown in figure 14.

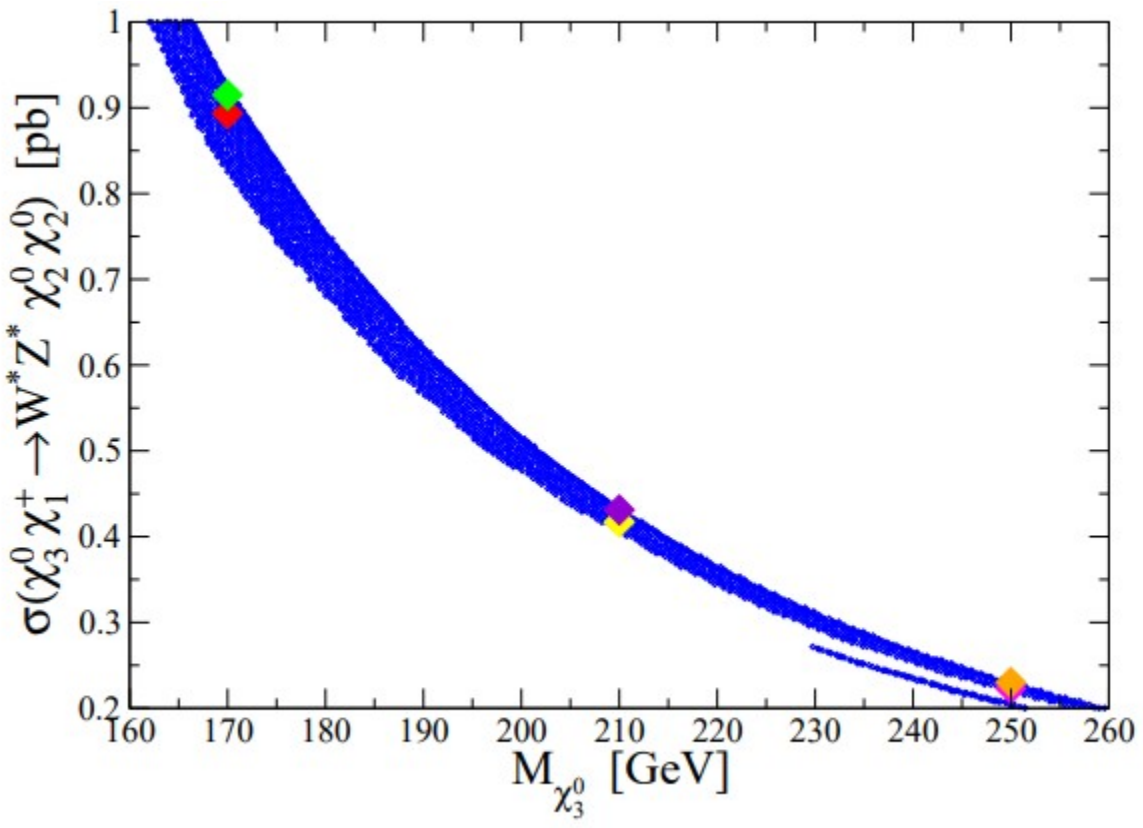

*Figure 14: Various solutions proposed by [23] to interpret, within NMSSM, the excesses observed at LHC.*

Assuming M1 ~μ, M2>>M1 and tanβ ~1, the cross section for **e+e- → χ+χ–** is ~300 fb for ECM=600 GeV and mχ+=250 GeV, assuming that the sneutrino exchange effect can be neglected.

ILC would collect $10^6$ such events, allowing a precise analysis of this channel.